\documentclass[reprint,nofootinbib,amsmath,amssymb,aps,eqsecnum,showpacs,twocolumn,superscriptaddress,prb,amsart]{revtex4-2} 
\usepackage{graphicx}
\usepackage{multirow}
\usepackage{color}
\usepackage{bm}
\usepackage{xcolor}
\usepackage{times}
\usepackage{amsmath,bm,amsfonts}
\usepackage{dcolumn}
\usepackage{graphicx}
\usepackage{latexsym}
\usepackage{hhline}
\usepackage{ulem}
\usepackage{braket}
\usepackage{bbold}
\usepackage{wasysym}
\usepackage{color}
\usepackage{xcolor}
\usepackage{hyperref}
\usepackage{comment}
\DeclareMathOperator{\tr}{tr}

\newcommand{\beq}{\begin{equation}}
\newcommand{\eeq}{\end{equation}}
\newcommand{\beqn}{\begin{eqnarray}}
\newcommand{\eeqn}{\end{eqnarray}}

\newcommand{\cE}{ {\cal E} }

\newcommand{\calB}{\mathcal{B}}

\definecolor{darkcyan}{rgb}{0.0, 0.55, 0.55}

\begin{document}
\title{Intrinsic mixed-state SPT from modulated symmetries and hierarchical structure of anomaly}

\author{Yizhi You}
\affiliation{Department of Physics, Northeastern University, Boston, MA, 02115, USA}
\author{Masaki Oshikawa}
\affiliation{Institute for Solid State Physics, University of Tokyo, Kashiwa, Chiba 277-8581, Japan }

\date{\today}

\begin{abstract}
We introduce a class of intrinsic symmetry-protected topological mixed-state(mSPT) in open quantum systems that feature modulated symmetries, such as dipole and subsystem symmetries. Intriguingly, these mSPT phases cannot be realized as the ground states of a gapped Hamiltonian under thermal equilibrium. The microscopic form of the density matrix characterizing these intrinsic mixed-state SPT ensembles is constructed using solvable coupled-wire models that incorporate quenched disorder or quantum channels. A detailed comparison of the hierarchical structure of boundary anomalies in both pure and mixed states is presented, utilizing flux insertion and Laughlin's charge pumping arguments. Finally, we explore the salient features of boundary anomalies in the mixed-state ensemble, which can be detected through the Rényi-N correlation function of charged observables.
\end{abstract}

\maketitle

\section{Introduction} 

Significant progress has been made in simulating entangled quantum states of matter using Noisy Intermediate-Scale Quantum (NISQ)\cite{bharti2022noisy} technology. In these systems, decoherence can occur under various conditions, leading to recent studies on the behavior of exotic quantum states in open systems subject to decoherence. When a quantum state interacts with its environment, it is continuously probed and measured, causing the state to become entangled with the environment (represented by ancilla qubits). Such system-ancilla coupling transforms the initial pure quantum state into a mixed-state ensemble. 
Consider the preparation of a symmetry-protected topological (SPT) state within a synthetic platform\cite{satzinger2021realizing,semeghini2021probing,chen2023realizing,iqbal2023creation,iqbal2023topological,tantivasadakarn2023hierarchy,baumer2023efficient,foss2023experimental} for measurement-based quantum computing (MBQC) purposes\cite{raussendorf2007topological,eckstein2024robust,zhu2023nishimori}. A critical question is whether quantum correlations can survive under decoherence when our initial pure SPT state transforms into a mixed state\cite{lee2022decoding, LeeYouXu2022, zhu2023nishimori}? Moreover, how can we classify and identify symmetry-protected mixed states (mSPT) in an open system that is far from equilibrium\cite{coser2019classification,bao2023mixed,fan2023diagnostics,chen2023symmetry,chen2023separability,sang2023mixed,li2024replica,zhang2022strange,lee2022symmetry,albert2014symmetries,brown2016quantum,peres1996separability,lessa2024mixed,khindanov2024robust}?

Recently, extensive research has focused on the mixed state symmetry-protected topological (mSPT) phase (also dubbed as \textit{Average-SPT} phase in some literature)\cite{lee2022symmetry,ma2023topological,sohal2024noisy,dai2023steady,wang2023intrinsic,de2022symmetry,chen2024unconventional,ma2023average,hauser2023continuous,Antinucci2023,lee2024exact,lu2024disentangling,lessa2024strong,khindanov2024robust,chirame2024stable} in open quantum systems settings.
The exploration of SPT in open systems has diversified into multiple perspectives. Ref.~\cite{fan2024diagnostics,zhang2024quantum,ma2024symmetry,lee2022symmetry,sang2024stability,zhang2024quantum} explores the impact of decoherence effects on an SPT wavefunction under quantum channels. Key questions arise, such as whether the topological structure and conditional mutual information are preserved when an SPT state undergoes weak measurement into a mixed state. Additionally, Ref.~\cite{lee2022symmetry,xue2024tensor,guo2024locally} explores and identifies the many-body invariants and experimentally accessible indicators that can detect and distinguish various mSPTs.
Another popular topic is whether decoherence introduces new phenomena into the realm of quantum phases in open quantum systems out of equilibrium\cite{ma2023topological,wang2023intrinsic,sohal2024noisy}. Specifically, it examines whether quantum channels can create unique mixed-state SPTs that have no counterparts in thermal equilibrium (such as the ground state of a local Hamiltonian).
Ref.~\cite{ma2023topological,zhang2023fractonic} introduces the concept of intrinsic mixed-state SPTs, which emerge from implementing quenched disorder or quantum channels to an intrinsic gapless SPT state. This raises the question of whether a broader class of intrinsic mSPT exists and how they can be characterized.

In a parallel effort, an increasing amount of research has been dedicated to the study of SPT phases with conservation of modulated symmetries, including charge multipole symmetry and subsystem symmetries~\cite{sachdev02,seidel05,pretko17,pretko18,he2020lieb,you2018higher,you2019higher,you20,you2021multipolar,gromov20,Dubinkin-dipole,may2021crystalline,delfino20232d,seiberg22a,seiberg23,feldmeier,lake1,lake2,lake23,delfino2023anyon,oh22b,glorioso2023goldstone,huang2023chern,burnell2023filling,dipolar-SPT,sala2024exotic,lam23,ebisu2310,du2024noncommutative,du2023quantum,han2024dipolar,sala2022dynamics,pace2024}. Systems with modulated symmetries characteristically exhibit constrained dynamics for charged excitations, as these would inevitably violate the symmetry constraints~\cite{vijay2016fracton,pai2019fracton,nandkishore2019fractons,pretko2020fracton}. Ref.~\cite{lam2024topological} explores the scope of dipole symmetry-protected SPTs by constructing coupled-wire models in 2D, denoted as the \textit{topological dipole insulator} (TDI). These models possess conserved charge and conserved x-dipole moment, are incompressible in the bulk, and host localized gapless modes at their boundaries\cite{prem18,shirley2019foliated,you2020symmetric,sullivan2020fractonal,sullivan2021fractonic,burnell22,huang2023chern, sullivan2021weak,zhang2023classification,you2019higher,radzihovsky2022lifshitz,may2022interaction,zhang2023classification,han2024topological,bertolini2024hall}. The edge of the TDI is characterized as a quadrupolar channel that displays a dipole $U^d(1)$ anomaly. A quantized amount of dipole gets transferred between the edges under the dipolar flux insertion, manifesting as the `quantized quadrupolar Hall effect' for TDIs\cite{prem2017emergent,huang2023chern,lam2024topological}. In addition, it was delineated in Ref.~\cite{lam2024topological} that a self-anomaly related to the \(U^e(1)\) symmetry or a mixed anomaly between \(U^{d}(1)\) and \(U^e(1)\) cannot exist as the 1D boundary of a TDI.

In this work, we explore the scope of symmetry-protected topological mixed-state (mSPT) with modulated symmetry\cite{sala2022dynamics,sala2023exotic,sala2020ergodicity} in open quantum systems, revealing a fertile ground for generating intrinsic mSPT states that exist exclusively in mixed states.
Contrary to the naive intuition that decoherence might trivialize an SPT wave function and diminish its entanglement structure, our results suggest that quantum decoherence can enrich the landscape of SPT states\cite{ma2023topological,zhang2023fractonic}.
Notably, we demonstrate a class of intrinsic mixed-state SPTs with modulated symmetries (dipole and subsystem symmetry)\cite{zhang2023classification,zhang2023fractonic} that host novel edge patterns and boundary anomalies, which have no counterparts in thermal equilibrium. We build intrinsic mSPTs by constructing the exact form of the density matrix from coupled-wire models, incorporating quantum channels or quenched disorder.
A detailed comparison of the hierarchical structure of boundary anomalies in both pure and mixed states is presented using flux insertion and charge pumping arguments. Additionally, we demonstrate that the edge anomaly\cite{ma2024symmetry,lessa2024mixed} in the mixed-state ensemble can be detected through the regular and Renyi-2 correlation function of charged observables. Specifically, if the edge exhibits a perturbative anomaly between strong charge and weak dipole symmetry, the dipole-charged operator shows quasi-long-range order in the mixed ensemble, whereas the charge operator exhibits quasi-long-range order only in the Renyi-2 correlator. This result provides a feasible route for detecting mixed-state anomalies through numerical simulations or experimental setups\cite{zhang2023synthetic}.

The paper is organized as follows: In Sec.~\ref{sec:review}, we review the no-go theorem for 2D SPT with dipole conservation (denoted TDI) in closed systems and demonstrate how this theorem can be partially relaxed in open systems where dipole conservation becomes a weak symmetry. In Sec.~\ref{sec:dipole}, we introduce a coupled wire construction to realize these intrinsic TDI under quenched disorder and quantum channels. We analyze the resulting edge anomaly through quasi-long-range ordered Renyi-2 correlations and demonstrate the stability of such edge long-range order (LRO) in a mixed ensemble through purification arguments. 
Finally, in Sec.~\ref{sec:3dhoti}, we extend our exploration of intrinsic mixed-state SPTs to 3D and present an intrinsic higher-order topological insulator with weak subsystem charge conservation and strong global U(1) symmetry, which supports chiral modes at the hinges.

\section{Preliminary and Background}

\subsection{review of no-go theorem for 2D topological dipole insulators in closed systems}\label{sec:review}

Before we delve into our exploration of intrinsic 2D topological dipole insulators(TDI) in mixed-state ensembles, let's briefly review the no-go theorem and hierarchy structure for 2D topological dipole insulators(TDI) at zero temperature under thermal equilibrium.

Consider a 2D insulator with charge ($U^e(1)$) and x-dipole moment ($U^d(1)$) conservation:
\begin{align} Q = \int \rho (x,y) ~dx dy ,\quad Q[x] = \int x \rho (x,y)~ dx dy \end{align}
where $\rho(x,y)$ is the charge density. 
The conservation of the x-dipole moment, \(Q[x]\), imposes restrictions on the dynamics of single particles: a charged excitation is allowed to move along the y-direction, whereas movement along the x-direction requires a neutral entity, such as a dipole-bound state. Re~\cite{lam2024topological} introduces a novel category of gapped topological dipole insulators protected by \(U^e(1)\) and \(U^d(1)\) symmetry. These insulators feature boundaries that support gapless modes with quadrupolar channels, revealing a \textit{self-anomaly} related to the \textit{x-dipole moment}. Specifically, when the system is placed on an open cylinder along the x-direction and subjected to \(U^{d}(1)\) flux insertion, the local dipole moment at the left ($x=1$) and right ($x=L$) edges respectively increases and decreases. This results in the transfer of dipole moment between the two boundaries, similar to the Laughlin argument in the quantum Hall effect (QHE), where a net charge transfer occurs across the bulk under charge flux insertion. 

Ref.~\cite{lam2024topological} presents a no-go theorem through flux insertion argument that precludes the existence of a mixed anomaly between dipole (\(U^{d}(1)\)) and charge (\(U^{e}(1)\)) symmetries at the edge of any 2D dipole insulators with finite correlation length in the bulk. 
The theorem is based on the premise that such an anomalous edge configuration would inevitably make the entire 2D bulk theory anomalous, which cannot be manifested in a lattice model with local interaction\footnote{However, these bulk anomalies can appear in gapless systems (e.g., intrinsic gapless SPT) where the anomaly emerges at the infrared (IR) scale. In this work, we only focus on the cases where the bulk is short-range entangled with finite correlation length.}. Consider a scenario where there is a mixed anomaly between the \(U^{d}(1)\) and \(U^{e}(1)\) symmetries at the edges, and we place the system on a cylinder along the x-direction. In this case, the insertion of a \(U^{d}(1)\) flux along the y-cycle, implemented by adding a gauge potential $A_y=\frac{2\pi x}{L_y}$, would lead to a change in charge density at the left boundary, at \(x=1\), increasing by \(q\). To maintain the conservation of total charge throughout the system, the charge density at the right edge, at \(x=L\), must decrease by \(q\). 
Consequently, the insertion of dipole flux results in the transfer of charge between the two boundaries. However, such a charge transfer between the edges also alters the bulk dipole moment of the entire 2D system and immediately violates the conservation of the dipole moment in the bulk, rendering the whole 2D theory anomalous under \(U^{d}(1)\) symmetry. Therefore, a mixed anomaly between \(U^{d}(1)\) and \(U^{e}(1)\) at the edge of any 2D insulators is not possible. Building on the same principles, it can be shown that the \(U^{e}(1)\) anomaly cannot manifest at the boundary because inserting a charge flux would induce charge pumping and alter the bulk dipole moment of the entire 2D system.

This observation leads to a broader principle: \textit{the hierarchical structure of quantum anomalies} in systems that conserve the charge of higher moments. Consider, for instance, a gapped bulk that conserves the $n-th$ order multipole moment, such as the octupole moment with \(n=3\). In this circumstance, the only feasible anomaly at the boundary is a self-anomaly for the charge octupole. If the boundary were to exhibit a mixed anomaly involving both the charge octupole and charge quadrupole, then inserting a flux corresponding to the octupole symmetry would initiate the pumping of quadrupoles between the edges, consequently altering the bulk octupole moment. This scenario suggests that the presence of a mixed anomaly between octupole and quadrupole symmetries at the boundary implies a concurrent self-anomaly for the octupole symmetry in the bulk. 

This phenomenon reflects a hierarchical structure of quantum anomalies in systems with multipole symmetry. The foundation for this hierarchical structure is that the symmetry and conservation laws governing different multipole moments are interdependent.
In a dipole-conserved system, the total charge density must be both conserved and neutral. Notably, the dipole symmetry \(U^d(1)\) cannot be treated as an internal symmetry, as it undergoes nontrivial transformations under spatial translation as \(T_x: Q[x] \rightarrow Q[x] + Q\). Namely, the lattice translation acting on the dipole operator \(Q[x]\) generates additional charge density \(Q\). Consequently, inducing a change in charge density at different positions along the x-axis results in distinct dipole moments. This scenario implies that a mixed anomaly between \(U^d(1)\) and \(U^e(1)\) at the left boundary presents a challenge, as it is impossible to consistently allocate an opposing anomaly pattern on the right boundary to neutralize the one on the left.

\subsection{Strong and weak symmetry in mixed states}

Symmetry plays a crucial role in understanding the complexities of highly entangled quantum states, and its interplay can be even richer in the context of mixed quantum states. When dealing with a mixed state described by a density matrix, there are two distinct types of symmetries: \textit{strong} and \textit{weak} symmetry. \textit{Weak symmetry} occurs when the density matrix remains invariant under a symmetry transformation \( U_g \), acting on both the left (ket) and right (bra) sides of the density matrix, as \(\rho = U_g \rho U_g^*\). Physically, this requires the density matrix to be block-diagonal, with each block corresponding to a different charge under \(G\).
In contrast, \textit{strong symmetry} for the density matrix demands invariance under a symmetry transformation represented by \(\rho=e^{i\theta}U_g \rho\), which acts solely on either the left (ket) or the right (bra) part of the density matrix, where \(e^{i\theta}\) is a global phase. The strong symmetry condition requires that all eigenstates of the density matrix also be eigenstates of the symmetry \(G\), each carrying the same charge.

\section{Approaching from an open system view: imSPT and modulated symmetry} \label{sec:dipole}

Open quantum systems provide a novel platform for investigating dynamical quantum many-body phases far from equilibrium. Ref.\cite{ma2023topological} introduces a new category of \textit{intrinsic mixed-state symmetry-protected topological (imSPT) phases} driven by disorder or quantum channels, which lack counterparts in equilibrium systems. These imSPT phases in open quantum systems can trace their origins to `intrinsic gapless SPT' Ref.\cite{thorngren2021intrinsically, li2023intrinsically, wen2023bulk} in closed systems, where the low-energy effective theory exhibits emergent anomalies in the bulk with a gapless spectrum. By incorporating either quenched disorder or dynamic quantum channels, the initial gapless pure state evolves into a short-range entangled mixed state, exemplifying the imSPT phase. We aim to explore alternative pathways for realizing intrinsic symmetry-protected topological mixed-states (imSPT) in systems featuring modulated charge conservation, such as those with dipole symmetry or subsystem symmetry. Specifically, we will demonstrate that open quantum systems support a broader class of mixed-state SPTs with modulated charge conservation, unique to open systems. Notably, the hierarchical structure of quantum anomalies in dipole-conserved systems discussed in Sec.~\ref{sec:review} will be modified in mixed states.

\subsection{Intrinsic Topological dipole insulator in open quantum systems}

We begin by detailing a microscopic construction for an intrinsic topological dipole insulator (TDI) in open quantum systems, which exhibits a mixed anomaly between \(U^{d}(1)\) and \(U^e(1)\) at the boundary. Although such a boundary anomaly cannot exist as the ground state of a gapped Hamiltonian due to the hierarchical anomaly structure imposed by the aforementioned no-go theorem, an open system described by a mixed-state density matrix offers a novel path to bypass this limitation.
Specifically, we will explore the topological dipole insulator in mixed states that features \textit{strong \(U^{e}(1)\) symmetry} and \textit{weak \(U^{d}(1)\) symmetry}. In this context, there is no \(U^{e}(1)\) charge exchange between the system and the environment (denoted as the ancilla). The system alone conserves its own $U^e(1)$ charge, while the dipole moment can fluctuate between the system and the ancilla.
By reducing the dipole symmetry to a \textit{weak symmetry} in the mixed state, which acts on both the ket and bra spaces of the density matrix, the aforementioned bulk anomaly associated with \(U^{d}(1)\) symmetry would be diminished. We will revisit this point later.

To construct a mixed-state ensemble that manifests an intrinsic topological dipole insulator, we will follow two steps:

1) We begin with a solvable microscopic Hamiltonian in the closed system using coupled wire constructions. Notably, the coupled wire approach will only gap out partial degrees of freedom, leaving the bulk gapless.

2) We consider the decoherence effect by adding quantum channels to the gapless state. Notably, the quantum channels we consider in this paper are not necessarily of finite depth, as we are focusing on the non-perturbative effects in systems with strong correlations. An infinite-depth quantum channel can be viewed as an infinite-depth unitary gate operating on the system and ancilla, which makes them strongly coupled. 
 Alternatively, one can also introduce quenched disorders instead of quantum channels, resulting in the system being characterized by a mixed-state disorder ensemble.
We will demonstrate that the resultant mixed-state density matrix is short-range correlated in the bulk and exhibits a mixed anomaly between \textit{strong \(U^{e}(1)\) symmetry} and \textit{weak \(U^{d}(1)\) symmetry} at the boundary.

\subsubsection{Step I: Coupled wire approach}\label{sec:coupledwire}

We first follow the footsteps of coupled wire constructions, which have been instrumental in developing various topological models with gapless boundaries \cite{poilblanc1987quantized, teo2014luttinger, vazifeh2013weyl, meng2015fractional, sagi2015array, iadecola2016wire, zhang2022coupledwire}. For instance, the coupled-wire construction of the integer quantum Hall system places a pair of \(L\)- and \(R\)-moving Luttinger liquids within each cell and introduces an inter-cell coupling term. This term effectively gaps out all channels except for the two outermost edge channels.
Adapting this approach to our scenario, we consider a 2D lattice that spans the \(xy\) plane, composed of 1D wires extending in the \(y\)-direction. For clarity, each unit cell contains two flavors of 1D charged Luttinger liquid per wire. We label these modes as \(\phi^a_{L/R}(r)\) (\(a=1,2\)).

\begin{figure}[h]
    \centering
\includegraphics[width=0.5\textwidth]{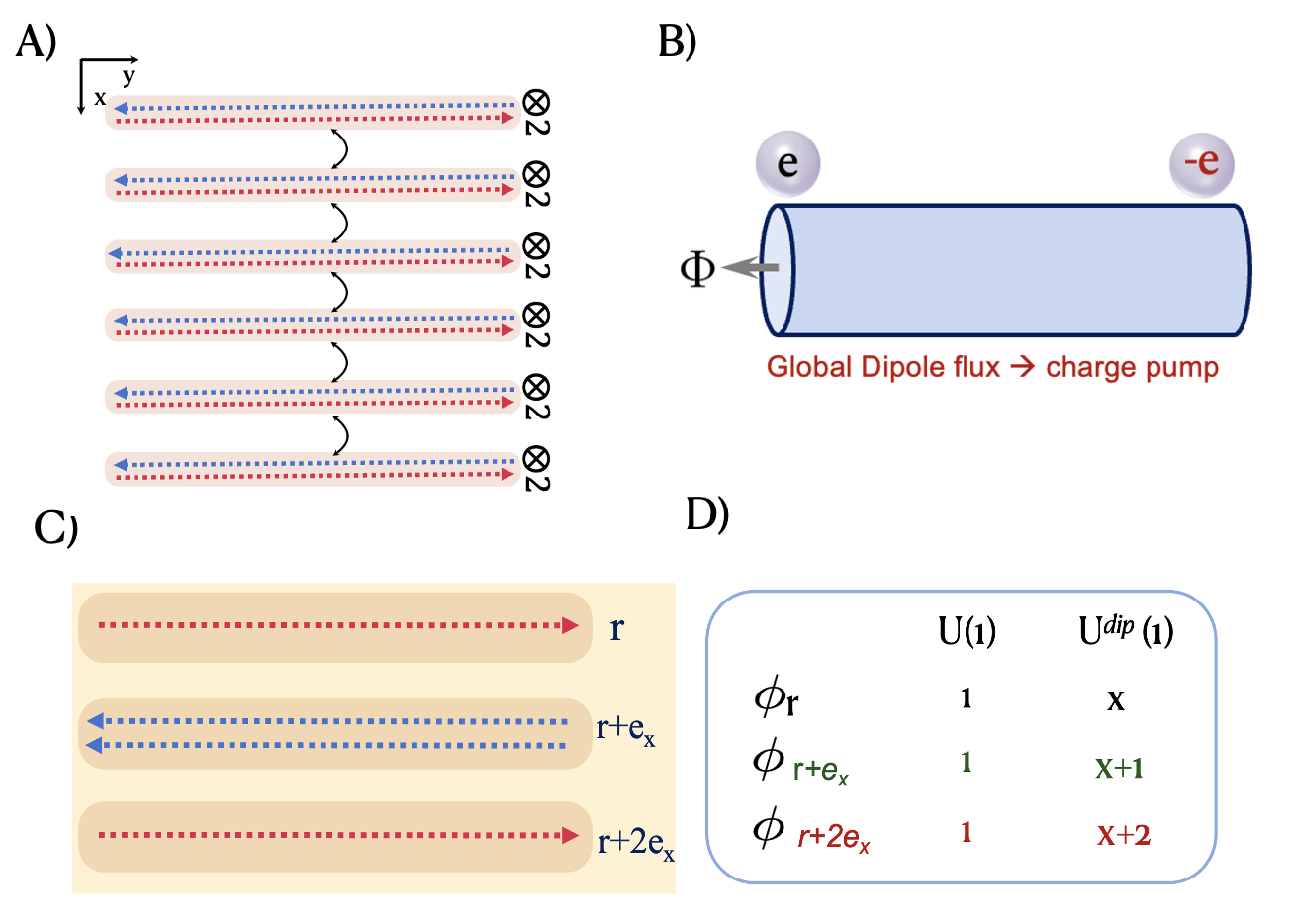}
    \caption{A) Coupled wire setting for TDI: each row represents two flavors of $1$D Dirac fermions of either $R$ (blue) or $L$ (red) chirality. B) Inserting a dipole flux through the cylinder triggers charge pumping between the edges. C) The building block of the coupled-wire construction. D) How the charges in each wire transform under the $U^e(1)$ and $U^d(1)$ symmetries.}
    \label{dipole}
\end{figure}

To streamline the formulation, we select a set of chiral modes from several adjacent rows as fundamental units (denoted as the \textit{building blocks}) and limit the inter-wire coupling to within these building blocks.
In our construction,
the building block $\calB_i$ stretches from the \(i\)-th to the \(i+2\)-th row,
and include the two left-moving modes \(\phi^{1}_{L}(r)\) and \(\phi^{2}_{L}(r+2e_x)\),
and two right-moving modes \(\phi^{1,2}_{R}(r+e_x)\).
While it spatially overlaps with the neighboring building blocks \(\calB_{i-2}, \calB_{i-1}, \calB_{i+1}\), and \(\calB_{i+2}\) (except at the boundaries), each chiral mode belongs to a unique building block. Thus, all the chiral modes of the system are decomposed into non-overlapped building blocks, and there is no interaction among different building blocks. It should be noted that the left-moving and right-moving chiral modes on the same row are assigned to different building blocks.


Each building block encompasses two helical boson modes, necessitating two independent mass terms to gap them out. The chiral boson fields transform under charge \(U^e(1)\) and dipole \(U^d(1)\) symmetry as follows:
\begin{align}
&U^d(1):\phi^a_{L/R}(x,y) \rightarrow \phi^a_{L/R}(x,y)+ \beta \cdot x \nonumber\\
&U^e(1): \phi^a_{L/R}(x,y) \rightarrow \phi^a_{L/R}(x,y)+ \alpha 
\label{eq:dipoleDefU(1)}
\end{align}
Notably, the $U^d(1)$ symmetry shifts the chiral boson field in a manner that depends on its x-position.


Let us introduce interactions within each building block such that they respect both \( U^e(1) \) and \( U^d(1) \) symmetries. For later convenience, we define the new fields in each building block as follows:
\begin{align}
&2\phi^A: \phi^1_{L,\bm{r}}-\phi^2_{R,\bm{r+e_x}}-
\phi^1_{R,\bm{r+e_x}}+\phi^2_{L,\bm{r+2e_x}} \nonumber\\
&2\phi^B: \phi^1_{L,\bm{r}}-\phi^2_{R,\bm{r+e_x}}+\phi^1_{R,\bm{r+e_x}}-\phi^2_{L,\bm{r+2e_x}}\nonumber\\
&2\theta^A: \phi^1_{L,\bm{r}}+\phi^2_{R,\bm{r+e_x}}+\phi^1_{R,\bm{r+e_x}}+\phi^1_{L,\bm{r+2e_x}}\nonumber\\
&2\theta^B: \phi^1_{L,\bm{r}}+\phi^2_{R,\bm{r+e_x}}-\phi^1_{R,\bm{r+e_x}}-\phi^1_{L,\bm{r+2e_x}} \nonumber
\end{align}

In terms of these fields, the symmetries act as:
\begin{align}
&U^e(1):
\begin{cases}
     \phi^A \rightarrow \phi^A , & \theta^A \rightarrow \theta^A + \alpha  \\
     \phi^B \rightarrow \phi^B, & \theta^B \rightarrow \theta^B 
\end{cases} , \\
&U^d(1):
\begin{cases}
    \phi^A \rightarrow \phi^A , & \theta^A \rightarrow \theta^A +2\beta  \\
    \phi^B \rightarrow \phi^B -\beta, & \theta^B \rightarrow \theta^B - \beta 
\end{cases}
\end{align}

We find that \( \phi^A \) is neutral under charge and dipole symmetries, so we can add the following term to gap out the partial degree of freedom:
\begin{align}
V_{int} = v_0 \cos{\phi^A} .
\label{eq:BosonLagrangian}
\end{align}
In the meantime, all other terms are forbidden as they would inevitably break either \( U^e(1) \) or \( U^d(1) \) symmetry. Since each building block contains two gapless modes, a strong interaction term in Eq.~\ref{eq:BosonLagrangian} can gap one of the gapless branches (\( \phi^A \)), leaving the \( \phi^B \) mode gapless. Notably, both \( \phi^B \) and its conjugate variable \( \theta^B \) are charged under \( U^d(1) \). This condition implies that the gapless branch of \( \phi^B \) displays a 't Hooft anomaly under \( U^d(1) \) symmetry, making it impossible to symmetrically gap out \( \phi^B \) in each building block in a closed system.


\subsubsection{disorder and decoherence}

We now consider integrating the building block into open quantum system settings by introducing quenched disorder or decoherence quantum channels \textit{within each building block}. This approach effectively transforms the gapless state, which has a divergent correlation length, in each building block into a mixed-state ensemble with short-range correlations.
To simplify the notation, we express the chiral boson fields in the coupled wire construction in terms of chiral fermion operators as \(\psi_{R/L} \sim e^{i\phi_{R/L}}\). In this representation, each building block is equivalent to a four-component Dirac spinor coupled to an O(4) dynamical mass vector:
\begin{align}\label{diracmass}
    &H=\psi^{\dagger}(k_y \sigma^{33}+m_1 \sigma^{13}+m_2 \sigma^{23}+m_3 \sigma^{01}+m_4 \sigma^{02})\psi  \nonumber\\
    &\psi^{\dagger}=(\psi^{\dagger}_{L,1},\psi^{\dagger}_{R,2},\psi^{\dagger}_{R,3},\psi^{\dagger}_{L,4}),\nonumber\\
    &\psi_{L,1}=e^{ i \phi^1_{L,\bm{r}}},
       \psi_{R,2}=e^{ i \phi^1_{R,\bm{r+e_x}}},\nonumber\\
          &\psi_{R,3}=e^{ i \phi^2_{R,\bm{r+e_x}}},
             \psi_{L,4}=e^{ i \phi^2_{L,\bm{r+2e_x}}}
\end{align}
Here, and throughout, we use the shorthand \(\sigma^{ij} \equiv \sigma^i \otimes \sigma^j \), where \(\sigma^i\) are the Pauli matrices.
The four components of the Dirac spinor represent the four chiral bosons in each building block, spread over the \(i\)-th to \(i+2\)-th wires. The dipole symmetry acts on the Dirac spinor as follows:
\begin{align}
     U^d(1): &   (\psi^{\dagger}_{L,1},\psi^{\dagger}_{R,2},\psi^{\dagger}_{R,3},\psi^{\dagger}_{L,4})\nonumber\\ &\rightarrow    (\psi^{\dagger}_{L,1},e^{i\beta}\psi^{\dagger}_{R,2},e^{i\beta}\psi^{\dagger}_{R,3},e^{i2\beta}\psi^{\dagger}_{L,4}) 
\end{align}
The fermion mass vectors \(\vec{m}=(m_1, m_2, m_3, m_4)\) in Eq.~\ref{diracmass} explicitly break dipole symmetry. If we decompose the O(4) vector into two O(2) bosons,
\( (\vec{q}_a,\vec{q}_b) = \left( m_1 + i m_2 , m_3+im_4 \right) \),
the \(U^d(1)\) transforms the dynamical mass as follows:
\begin{align}
     U^d(1): &  (\vec{q}_a,\vec{q}_b) \rightarrow (e^{i\beta}\vec{q}_a,e^{i\beta}\vec{q}_b).
\end{align}
Thus, they cannot manifest as a static fermion bilinear mass. However, we can consider them as a dynamical mass\cite{bi2015classification,you2014symmetry,you2015bridging,you2016quantum} that fluctuates in space-time without establishing order, thereby preserving dipole symmetry. These dynamical mass vectors can be generated through the dipole-neutral
interaction term specified in Eq.~\ref{eq:BosonLagrangian}.
In the fermion representation, the inter-wire interaction is transformed into a fermion quartet term, \(\psi^{\dagger}_{L,1}\psi_{R,2}\psi_{R,3}\psi^{\dagger}_{L,4} + h.c.\). The Hubbard-Stratonovich transformation of this term indeed produces the dynamic O(4) vectors \(\vec{m}=(m_1, m_2, m_3, m_4)\), which are coupled to the fermion bilinears as described in Eq.~\ref{diracmass}.

One can further integrate the fermions in each building block to obtain the effective theory for the dynamic mass \(\vec{m}\):
\begin{equation}\begin{split}
&\mathcal{L}=\frac{1}{g}(\partial_{\mu} \vec{m})^2+\frac{2\pi}{\Omega^3} \int_0^1 du\epsilon^{ijkl}  m_{i}\partial_y m_{j} \partial_t m_{k}\partial_u m_{l},\\
 &\vec{m}(y,t,u=0)=(1,0,0,0),\phantom{=}
 \vec{m}(y,t,u=1)=\vec{m}(y,t),
\end{split}\label{eq:wzw}
\end{equation}
The effective theory for the dynamical mass vector resembles a nonlinear sigma model (NLSM) with an \(O(4)_1\) Wess-Zumino-Witten (WZW) term\cite{xu2013nonperturbative}. In this scenario, the fermionic excitations in each building block are gapped through symmetric mass generation\cite{you2014symmetry,you2015bridging,you2016quantum,xu2021green,you2018symmetric,wang2022symmetric}, while a gapless collective bosonic excitation arises from the fluctuations of the vector field \(\vec{m}=(m_1, m_2, m_3, m_4)\). 
It is worth emphasizing that the WZW term for the dynamical mass is crucial for maintaining the gapless nature of each quasi-1D building block in a closed system. It reflects the perturbative anomaly for the \(U^d(1)\) symmetry\cite{xu2013wave,bi2015classification,you2014wave}, indicating that there is no way to gap out the mass vectors unless dipole conservation is broken. Moreover, the WZW term induces a Berry phase for the dynamical mass fluctuations, preventing the O(4) vector from becoming disordered with finite correlation lengths.

Now consider adding quenched disorder to each building block by introducing the random disordered mass vector \(\vec{m}(r,t)\) in Eq.~\ref{diracmass}. Although each specific disorder mass pattern explicitly breaks dipole conservation, when we consider mixed ensembles of all possible disorder mass configurations, the resulting mixed-state density matrix \(\rho\) still exhibits a weak \(U^{d}(1)\) symmetry. The mixed-state density matrix in each building block can be expressed as:
\begin{align}\label{mixed0}
    \rho=\sum_{\{\vec{m} \}} |\vec{m}\rangle \langle \vec{m} |
\end{align}
For each specific pattern \(\vec{m}\), the ket vector \(|\vec{m}\rangle\) represents the ground state of the 1D Dirac spinor in Eq.~\ref{diracmass}, coupled to the \textit{static vector mass} \(\vec{m}\) in each building block, resulting in a gapped, short-range correlated state. The density matrix in Eq.~\ref{mixed0} comprises a convex sum of 1D gapped fermions with a disordered vector mass \(\vec{m} = (m_1, m_2, m_3, m_4)\). Although the \(|\vec{m}\rangle\) state breaks the dipole symmetry, the incoherent sum of all possible patterns of \(|\vec{m}\rangle\) still exhibits weak dipole symmetry.
Notably, the WZW term in Eq.~\ref{eq:wzw} vanishes in the mixed state \(\rho\), as its effect is nullified by the opposite Berry phase from the bra and ket spaces.

Alternatively, one can implement quantum channels to transform the gapless pure state in each building block into a short-range correlated mixed ensemble. To clarify, implementing a quantum channel with infinite depth is essential for driving the mixed state toward a short-range entangled state. A finite-depth quantum channel (FDQC) is equivalent to a finite-depth unitary acting on the purified state, in the presence of additional ancillae. Therefore, an FDQC may not drive the purified state into a gapped bulk with a finite correlation length, especially given our initial state has a gapless building block.

We use the Choi-Jamiolkowski isomorphism to express the decoherence effect of each building block. As demonstrated in Sec.~\ref{sec:coupledwire}, without decoherence or disorder, the building blocks contain a gapless boson \( \phi^B \) and its conjugate partner \( \theta^B \), both charged under \( U^d(1) \). Considering an open quantum system characterized by a mixed-state density matrix, we can express the effective theory of each building block in the Choi-doubled Hilbert space, which is composed of two copies of the gapless bosons from the ket and bra spaces:
\begin{align}
\mathcal{S}_{\text{Choi-doubled}}&=\mathcal{S}^l[\theta^B_l,\phi^B_l]-\mathcal{S}^r[\theta^B_r,\phi^B_r]\nonumber\\
&+\mu \cos(\theta^B_l-\phi^B_r)+\mu \cos(\phi^B_l-\theta^B_r)\nonumber\\
\mathcal{S}[\theta^B,\phi^B]&=\int\mathrm{d}x\mathrm{d}\tau\frac{1}{4\pi}\left(\partial_x\theta^B\partial_\tau\phi^B+\partial_x\phi^B\partial_\tau\theta^B\right)\nonumber\\
&-\frac{1}{8\pi}\left[K_0(\partial_x\theta^B)^2+\frac{4}{K_0}(\partial_x\phi^B)^2\right],\nonumber\\
U^d(1): ~&
    \phi^B_{l/r} \rightarrow \phi^B_{l/r} -\beta, ~\theta^B_{l/r} \rightarrow \theta^B_{l/r} - \beta 
\label{choidouble}
\end{align}
The terms \(\theta^B_{l/r}\) and \(\phi^B_{l/r}\) refer to the boson field in the ket/bra space within the Euclidean space path integral, each representing a gapless Luttinger liquid in the ket/bra space.
The coupling term between the \(l/r\) modes is invariant under weak dipole symmetry.
As analyzed in Ref.~\cite{zhang2023fractonic}, for sufficiently large \(\mu\), the two gapless bosons in the Choi-doubled space can be fully gapped, resulting in a mixed state with short-range correlations. Consequently, the quantum channels can transform the gapless building block into a short-range entangled mixed state.

\subsection{Edge anomaly and correlation for mixed state ensembles}\label{sec:dipoleanomaly}

To this end, we have demonstrated that inter-wire interactions combined with quenched disorders render all building blocks into a mixed state with short-range correlations. Therefore, the bulk degrees of freedom achieve a short-range entangled mixed state. But what happens at the boundary?
From the building block perspective, we identify a set of isolated modes \(\phi^1_{R}(0,y), \phi^2_{L}(0,y), \phi^2_{R}(0,y), \phi^2_{L}(1,y)\) at the left boundary, spanning from the row at \(x=0\) to \(x=1\) (for simplicity, we will omit the y-coordinate in future discussions about the edge).
These modes remain decoupled from the rest of the building blocks.
Adding intra-wire coupling \(\cos(\phi^2_{L}(0) - \phi^2_{R}(0))\) would partially gap out one branch while leaving the remaining helical branch \(\phi^1_{R}(0), \phi^2_{L}(1)\) gapless. 

The edge modes transform under charge and dipole symmetry as:
\begin{align}
&U^e(1):(\phi^1_{R}(0),\phi^2_{L}(1))\rightarrow (\phi^1_{R}(0)+\alpha,\phi^2_{L}(1)+\alpha)\nonumber\\
&U^d(1):(\phi^1_{R}(0),\phi^2_{L}(1))\rightarrow (\phi^1_{R}(0),\phi^2_{L}(1)+\beta) 
\end{align}
The dipole symmetry functions like a `chiral symmetry', introducing different charges between the left $\phi^2_{L}(1)$ and right $\phi^1_{R}(0)$ movers. Consequently, the edge theory exhibits a mixed anomaly between the charge $U^e(1)$ and dipole $U^d(1)$ symmetries, making it impossible to gap these modes in a closed system.

Under open system settings, we now consider a mixed ensemble with weak dipole \( U^d(1) \) symmetry. We introduce either quenched disorder or quantum channels to decohere the edge. The question arises whether it is possible to `trivialize the edge' in a mixed state while respecting both weak \( U^d(1) \) and strong \( U^e(1) \) symmetries. A few comments are in order:

1. When considering quantum channels, we will take into account the infinite-depth quantum channels and other non-perturbative effects as well. This approach is consistent with how we treat gapless edges in closed systems under thermal equilibrium, where it is necessary to demonstrate that the edge cannot be symmetrically short-range correlated under any strong interaction, rather than relying on perturbations that can be manifested by a local unitary circuit.

2. The concept of `trivializing the edge' in a mixed-state ensemble: In a closed system, an anomalous edge is typically identified by its energy spectrum or correlation function. In a mixed state, if an anomalous edge undergoes purification, it will yield an anomalous edge of a pure state \cite{lessa2024strong,lessa2024mixed,wang2024anomaly}. This implies that an anomalous mixed state \textit{cannot} be purified into a short-range entangled state with on-site symmetry actions.


To facilitate the discussion regarding mixed anomalies, we represent the helical edge modes \(\phi^1_{R}(0), \phi^2_{L}(1)\) at the left boundary using a two-component Dirac spinor coupling with dynamical masses:
\begin{align}\label{chiral}
     &\mathcal{H}=   \bm{\psi}^{\dagger}( i\partial_y \tau^{z} +m_1 \tau^{x} +m_2\tau^{y} )\bm{\psi}\nonumber\\
     &U^e(1):\bm{\psi}\rightarrow e^{i \alpha}\bm{\psi} \nonumber\\
&U^d(1):\bm{\psi}\rightarrow e^{i \beta \tau^z}\bm{\psi}
\end{align}
The chirality index is denoted by \(\tau^{z}\), indicating the separation of left and right movers at the edge, extending from the row at \(x=0\) to \(x=1\). The dynamical masses \(m_1\) and \(m_2\) fluctuate in spacetime without ordering.
In this context, \(U^d(1)\) functions as a chiral symmetry for the Dirac fermion. By combining the two chiral masses into a complex field \(\bm{m} = m_1 + i m_2\), the \(U^d(1)\) symmetry action induces a phase shift on the complex mass field: \(\bm{m} \rightarrow e^{i \beta} \bm{m}\).  

The decoherence process can be described using Kraus operators in a pure measurement channel:
\begin{align}\label{decohere0}
 & \rho = \mathcal{E}[\rho_0], \hat{\rho}_0=|\Psi \rangle \langle \Psi|,\ \ \mathcal{E} = \prod_{\vec{r}} \cE_{\vec{r}}, \nonumber \\ &  \cE_{\vec{r}}[\rho_0] = \frac{1}{2} \rho_0 + \frac{1}{2}  O^{\dagger}_{\vec{r}} \rho_0 O_{\vec{r}}. 
\end{align}
Here, \(\mathcal{E}\) is given as the composition of local decoherence channels \(\mathcal{E}_{\vec{r}}\). $\Psi$ is the ground state of the massless Dirac fermion (with $\bm{m}=0$) in Eq.~\ref{chiral}.
The operators \(O_{\vec{r}}\) can be selected to be:
\begin{align}\label{kraus}
   & O=\bm{\psi}^{\dagger}(\tau^{x} +i\tau^{y} )\bm{\psi}
\end{align}
Such a quantum channel can be viewed as a measurement of the chiral mass field \(\bm{m} = m_1 + i m_2\) and involves averaging over all outcomes. It triggers the gapless Dirac fermion in Eq.~\ref{chiral} into a mixed-state density matrix that incoherently aggregates over different smooth \(\bm{m}\) patterns:
\begin{align}\label{MIX}
  \rho=\sum_{\{\bm{m}\}} p_{\{\bm{m}\}} |\{\bm{m}\} \rangle \langle \{\bm{m}\}|
\end{align}
Each mass pattern \(|\{\bm{m}\}\rangle\) represents the ground state wavefunction of a massive Dirac fermion described in Eq.~\ref{chiral} with a \textit{static mass} \(\bm{m}\). \(p_{\bm{m}}\) denotes the probability distribution for each pattern, depending on the initial state \(\Psi\). Note that the mass patterns \(|\{\bm{m}\}\rangle\) are mostly smooth, as the initial state \(\Psi\), which we begin with before measurement, is the ground state of a massless Dirac fermion with $k_y \rightarrow 0$.
The density matrix in Eq.~\ref{MIX} is invariant under the weak \(U^d(1)\) symmetry and strong \(U^e(1)\) symmetry, meaning all eigenvectors of the density matrix \(|\{\bm{m}\}\rangle\) possess the same number of \(U^e(1)\) charges. Now consider inserting a global flux with respect to \(U^d(1)\) symmetry. This can be achieved by creating a \(2\pi\) winding number in space on the complex mass field \(\bm{m}=e^{\frac{i 2\pi y}{L_y}}\). As the \(1d\) Dirac fermion manifests a chiral anomaly, such a flux insertion would trigger a shift in the total \(U^e(1)\) charge, leading all eigenvectors \(|\{\bm{m}\}\rangle\) in the density matrix in Eq.~\ref{MIX} to accrue additional charge. This indicates that the edge density matrix \(\rho\) exhibits a mixed anomaly between strong \(U^e(1)\) and weak \(U^d(1)\) symmetries.

Finally, we briefly discuss the case of adding quenched disorder. Suppose we introduce random mass configurations to the Dirac fermion in Eq.~\ref{chiral} to model quenched disorder at the edge. When considering quenched disorder patterns of \(|\{\bm{m}\}\rangle\), the density matrix becomes an incoherent sum of all possible \(|\{\bm{m}\}\rangle\) configurations, both smooth and rough. These patterns include \(|\{\bm{m}\}\rangle\) configurations with no winding number along the 1D chain (with PBC), as well as those with \(2\pi N\) winding numbers. Each \(|\{\bm{m}\}\rangle\) represents the ground state of a Dirac fermion with a chiral mass \(\bm{m}\). However, patterns with or without winding numbers will have different \(U^e(1)\) charges in the ground state, indicating that different \(|\{\bm{m}\}\rangle\) configurations in the mixed state possess different \(U^e(1)\) quantum numbers. This explicitly breaks the strong \(U^e(1)\) symmetry. This situation mirrors the conflict of symmetry observed in anomalous SPT edges, where gauging (adding winding number) one symmetry explicitly breaks another. Thus, in the presence of quenched disorder at the edge, the strong symmetry is broken. Likewise, we can consider a scenario where the mixed state is a combination of all possible \(|\{\bm{m}\}\rangle\) configurations with the same winding number \(2\pi M\). Although each \(|\{\bm{m}\}\rangle\) pattern differs locally, they share the same winding number under PBC. In this case, each \(|\{\bm{m}\}\rangle\) configuration possesses the same \(U^e(1)\) charge, resulting in a density matrix being strongly \(U^e(1)\) symmetric. However, the charged operator in the Renyi-2 correlator would exhibit non-vanishing LRO as a result of spontaneous strong-to-weak symmetry breaking, as discussed in Refs.~\cite{lee2022symmetry,ma2023average,sala2024spontaneous,lessa2024strong}. This aligns with the study of mixed state anomaly between strong and weak symmetry in Ref.~\cite{xu2024average}, where disordering the weak symmetry inevitably triggers strong-to-weak symmetry breaking.

\subsection{Signatures of quantum anomaly in mixed states}

So far, we have developed a mixed-state topological dipole insulator, protected by strong \(U^e(1)\) and weak \(U^d(1)\) symmetries. Its bulk renders a short-entangled mixed state, and its boundary displays quantum anomalies. A pertinent issue is the detection of these boundary anomalies in a mixed state under open system settings. In closed systems, for example, when considering an SPT wavefunction as the ground state of a local Hamiltonian, edge anomalies are revealed through a boundary spectrum that is either degenerate or gapless. Similarly, the entanglement spectra of the SPT states exhibit resemblances to this edge spectrum, facilitating the identification of edge anomalies through bulk entanglement properties.
Analyzing the boundary dynamics of a mixed-state SPT presents significant challenges as none of the aforementioned measures can be applied to an open system. In open quantum systems, far from equilibrium, the traditional concept of an energy spectrum no longer holds. Moreover, since we consider a mixed-state density matrix, the spectrum of the reduced density matrix incorporates classical contributions, such as the classical entropy of the mixed state that scales with the volume, rendering traditional entanglement entropy diagnostics ineffective.
In Refs.~\cite{lessa2024mixed,wang2024anomaly}, it has been demonstrated that the characteristics of anomalies in a mixed state can be identified using separability criteria if the anomaly is triggered solely by strong symmetry.

In this section, we examine quantum anomalies at the boundary of the topological dipole insulator in an open system from an alternative perspective. Specifically, we will demonstrate that the mixed anomaly between strong \(U^e(1)\) and weak \(U^d(1)\) symmetries at the boundary can be detected through the Renyi-N correlation function of the density matrix at the edge, which exhibits the following characteristics:
\begin{align}\label{edgecor}
    &\tr[\rho G^*(y') G(y)]=\frac{1}{|y-y'|^a}\nonumber\\
    &\frac{\tr[\rho S(y') S^*(y)\rho S(y) S^*(y') ]}{\tr[\rho^2]}=\frac{1}{|y-y'|^{\eta}}
\end{align}
Here, \(G(x)\) is an operator carrying dipole \(U^d(1)\) charge, while \(S(x)\) is an operator carrying \(U^e(1)\) charge. The dipole-charged correlation function exhibits quasi-long-range order in the mixed ensemble, while the charged correlation function demonstrates quasi-long-range order in the Rényi-2 correlator\cite{xu2024average}. Finally, we would like to note that our focus here is on cases where the strong \(U^e(1)\) and weak \(U^d(1)\) symmetries remain intact on the edge, without being broken either explicitly or spontaneously. Similar to the SPT boundary in thermal equilibrium, symmetry-breaking patterns can occur on the edge. If this happens, disordering the weak symmetry would lead to the breaking of the strong symmetry, either explicitly, as discussed in Sec.\ref{sec:dipoleanomaly}, or spontaneously, as demonstrated in Ref.\cite{xu2024average}.

\subsubsection{Purification view of anomaly}\label{sec:puriano}

To elucidate the structure of the edge correlation function in Eq.~\ref{edgecor}, we adopt a purification perspective by analyzing the mixed-state density matrix from its purified state, which is defined on the enlarged Hilbert space in the presence of additional ancillae from the environment. We first redefine the helical modes \(\phi^1_{R}(0)\) and \(\phi^2_{L}(1)\) at the system's boundary in terms of boson operators as \(\phi^1_{R}(0) + \phi^2_{L}(1) = 2\theta^s\) and \(\phi^1_{R}(0) - \phi^2_{L}(1) = 2\phi^s\). These bosons transformation under charge and dipole symmetry as follows \footnote{One can also shift the coordinate of \(x\) to make \(\theta^s\) neutral under dipole symmetry.}:
\begin{align}
&U^e(1):\theta^s \rightarrow \theta^s+\alpha \nonumber\\
&U^d(1): \phi^s \rightarrow \phi^s-\beta
\end{align}
Before the introduction of quantum channels or disorders, the edge represents a gapless state that displays mixed anomalies between charge and dipole symmetries. Consequently, both \(e^{i\theta^s}\) and \(e^{i\phi^s}\) demonstrate quasi-long-range order with algebraically decaying correlations. When quantum channels are activated in open systems, this effectively introduces interactions that entangle the system with the ancillae. The stability of long-range correlations (including quasi-long-range order) for the operators \(e^{i\theta^s}\) and \(e^{i\phi^s}\) hinges on whether interactions between the system and the ancilla, introduced by quantum channels, can diminish the long-range order of these operators.

A few observations concerning the symmetry constraints in the purification picture are in order. Given that the mixed state possesses strong $U^e(1)$ symmetry, there are no charge fluctuations or exchanges between the system and the ancilla in the purified state. Meanwhile, dipole fluctuations and exchanges with the ancilla are permissible, provided that the purified state, which encompasses both the system and the ancilla, remains invariant under \(U^d(1)\).

To diminish the long-range order of \(e^{i\theta^s}\) in the purified state, one needs to proliferate its conjugate partner \(e^{i\phi^s}\). However, proliferating \(e^{i\phi^s}\) alone is problematic as it carries a dipole charge. Doing so would explicitly break dipole conservation in the purified state (which includes both the system and the ancilla), thereby breaking the weak \(U^d(1)\) symmetry in the mixed-state ensemble of the system. To circumvent this issue, we can consider a bound state operator \(e^{i(\phi^s-\phi^a)}\), where \(\phi^a\) is a dipole-charged operator that acts only within the ancilla space. This bound state is neutral under dipole symmetry, and its proliferation leads to strong fluctuations and finite correlation for operator \(e^{i\theta^s}\). As a result, the quasi-long-range order of \(e^{i\theta^s}\) is unstable under quantum channels.

Then, what is the fate of \(e^{i\phi^s}\)? To diminish its long-range order in the purified state, one needs to proliferate its conjugate partner, \(e^{i\theta^s}\). However, proliferating \(e^{i\theta^s}\) alone is problematic because it carries a \(U^e(1)\) charge. One might consider a bound state operator \(e^{i(\theta^s)}O_a\), with \(O_a\) acting on the ancillae. However, since the \(U^e(1)\) charge is carried solely by the system's qubits, the operator \(e^{i(\theta^s) }O_a\) remains charged under \(U^e(1)\) symmetry, regardless of the choice of \(O_a\). Therefore, there is no way to diminish the quasi-long-range order of \(e^{i\phi^s}\) under symmetry constraints, and it remains stable under quantum channels.

A more systematic statement can be elucidated as follows\footnote{We acknowledge Chong Wang for bringing about this anomaly argument.}. From the purification perspective, the dipole moment density operator (which is charged under \(U^e(1)\)) can be expressed as \(\rho^d = \rho^d_s \otimes I_a + I_s \otimes \rho^d_a\), with the first operator (\(\rho^d_s, I_s\)) acting on the system and the second operator (\(\rho^d_a, I_a\)) acting on the ancilla. Similarly, the charge density operator (which is charged under \(U^d(1)\)), is defined as \(\rho^e = \rho^e_s \otimes I_a\). In the purified state, due to the mixed anomaly between charge and dipole, both \(\rho^e\) and \(\rho^d\) operators exhibit quasi-long-range order. This suggests that the operator \(\rho^e_s\) (charged under \(U^d(1)\)) has algebraic decay correlation.
However, since \(\rho^d\) includes the sum of the local dipole moment from the system and the ancilla, there is no guarantee that \(\rho^d_s\) has long-range order, as LRO can be contributed by the ancilla part \(\rho^d_a\).

\subsubsection{Renyi-2 correlator from purification}

\begin{figure}[h]
    \centering
\includegraphics[width=0.5\textwidth]{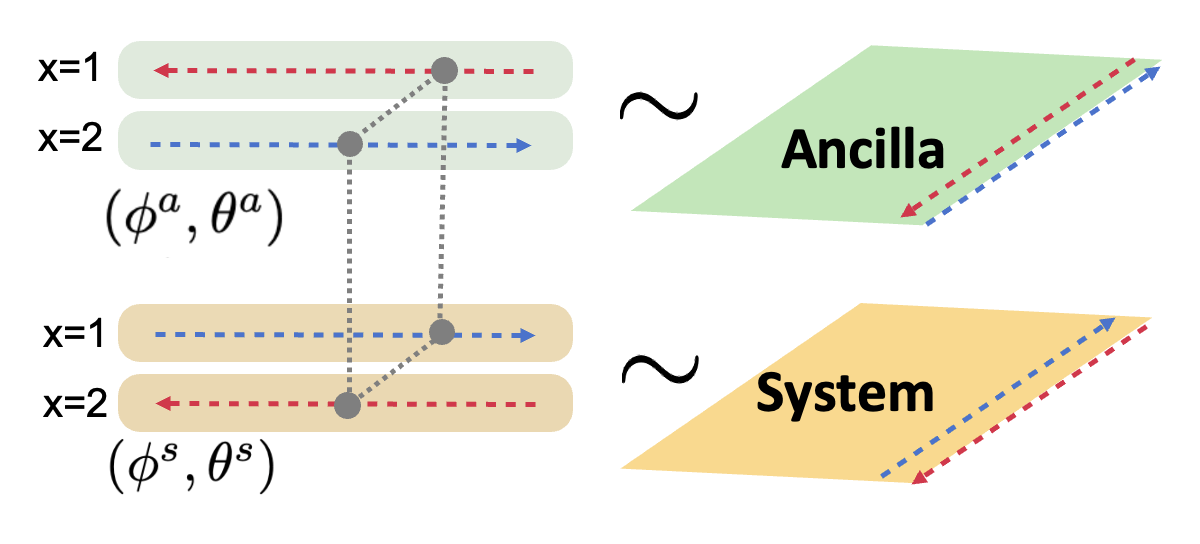}
    \caption{The purified state comprises the system (bottom layer) and the ancilla (top layer). The system's edge features a helical mode exhibiting a mixed anomaly between \(U^e(1)\) and \(U^d(1)\). Decoherence in the system layer occurs through entanglement with the ancilla.}
    \label{edge}
\end{figure}

While the charged operator $e^{i\theta^s}$ exhibits short-range correlation $\tr[\rho e^{i\theta^s(y)} e^{-i\theta^s(y')}] \rightarrow e^{-a|y-y'|}$, its Renyi-2 correlator still exhibits long-range order. To elucidate this, we introduce another set of helical modes from the ancilla's edge (denoted as $(\phi^a, \theta^a)$ with $[\phi^a(y), \theta^a(y')] = i\pi \Theta(y-y')$), which live atop the system's edge as Fig.~\ref{edge}. We can view this arrangement as having an ancilla layer on top of the system with the same coupled wire setting but opposite chirality, so the edge of the ancilla contains another set of helical modes illustrated as Fig.~\ref{edge}.

The system, along with the ancilla, transforms under symmetry as follows:
\begin{align}
&U^e(1):\theta^s \rightarrow \theta^s+\alpha \nonumber\\
&U^d(1):\phi^s \rightarrow \phi^s-\beta,~\phi^a \rightarrow \phi^a-\beta 
\end{align}
The \(U^e(1)\) charge is carried exclusively by the \(\theta^s\) field, so the system's density matrix \(\rho\), after tracing out the ancilla, exhibits strong \(U^e(1)\) symmetry. Meanwhile, both the \(\phi^s\) and \(\phi^a\) fields carry the x-dipole moment. A conserved dipole moment in the purified state shared by system and ancilla indicates that the system's density matrix \(\rho\) exhibits weak \(U^d(1)\) symmetry.
We can entangle the system with the ancilla in the following manner that respects both \(U^e(1)\) and \(U^d(1)\):
\begin{align}\label{edgecouple}
&H=g_1\cos(\phi^a-\phi^s)
\end{align}
In the regime of strong coupling, this interaction constrains the relative phase fluctuations between the system and the ancilla, setting \(\phi^a = \phi^s\). Thus, the relative phase between \(\phi^s\) and \(\phi^a\) is pinned. Consequently, Eq.~\ref{edgecouple} would gap out the \(\theta^a - \theta^s\) mode, while the symmetric combination of the system and ancilla, \(\theta^a + \theta^s\), remains gapless. Tracing out the ancilla from the purified state is analogous to measuring \(e^{i\phi^s}\) and summing over all possible outcomes.

If we measure the edge correlation function of the purified wavefunction \(|\psi^{as}\rangle\), the following modes are gapless and, hence, exhibit power-law correlation:
\begin{align}\label{purico}
& \langle\psi^{as}|e^{i(\theta^a(y)+\theta^s(y))} e^{-i(\theta^a(y')+\theta^s(y'))}|\psi^{as}\rangle=\frac{1}{|y-y'|^b}\nonumber\\
& \langle\psi^{as}|e^{i\phi^s(y)} e^{-i\phi^s(y')}|\psi^{as}\rangle=\frac{1}{|y-y'|^a}
\end{align}
The algebraic decay exponent in Eq.~\ref{purico} is influenced by the microscopic details of the interaction and is not universal. The quasi-long-range order of these operators is protected by the mixed anomaly between \(U^e(1)\) and \(U^d(1)\) symmetries. Specifically, the operator \(e^{i(\theta^a+\theta^s)}\) carries \(U^e(1)\) charge, while the operator \(e^{i \phi^s}\) carries the \(U^d(1)\) dipole moment. Since these two operators form canonical conjugate pairs, it is impossible to gap them out without breaking either \(U^e(1)\) or \(U^d(1)\) symmetry. Consequently, the edge of the purified state remains gapless, characterized by a mixed anomaly between \(U^e(1)\) and \(U^d(1)\) symmetries.

Once we trace out the ancilla degree of freedom from the purified wave function, the second part of Eq.~\ref{purico} reduces to the correlation function of \(e^{i\phi^s}\) in the mixed state:
\begin{align}
&  \tr[\rho e^{i \phi^s(y)} e^{-i \phi^s(y')}]=\frac{1}{|y-y'|^a}
\end{align}
The first part of Eq.~\ref{purico} can be tricky since the charged operator \(e^{i(\theta^a+\theta^s)}\) involves both the system and the ancilla. Once we trace out the ancilla, the operator acting on the ancilla, \(\theta^a\), becomes inaccessible. The only measurable property in the mixed-state density matrix is \(\theta^s\).
\begin{align}
&  \tr[\rho e^{i\theta^s(y)} e^{-i\theta^s(y')}]=\langle\psi^{as}|e^{i\theta^s(y)} e^{-i\theta^s(y')}|\psi^{as}\rangle \nonumber\\
&=\langle e^{i(\theta^a(y)+\theta^s(y))/2} e^{-i(\theta^a(y')+\theta^s(y'))/2} \nonumber\\
&e^{i(-\theta^a(y)+\theta^s(y))/2} e^{-i(-\theta^a(y')+\theta^s(y'))/2}\rangle\nonumber\\
&\sim \langle e^{i(\theta^a(y)+\theta^s(y))/2} e^{-i(\theta^a(y')+\theta^s(y'))/2}\rangle \nonumber\\
&\langle e^{i(-\theta^a(y)+\theta^s(y))/2} e^{-i(-\theta^a(y')+\theta^s(y'))/2}\rangle
\rightarrow e^{-\lambda |y-y'|}
\end{align}
In the final step, we leverage the fact that the mode \(\theta^a-\theta^s\) is gapped, resulting in \(\theta^a\) exhibiting short-range correlation. This observation aligns with insights from the purified wavefunction. Although the \(\theta^a+\theta^s\) mode exhibits long-range correlation, tracing out the ancilla effectively averages out the strong fluctuations of \(\theta^a\), thereby diminishing the long-range order of \(\theta^s\). Nonetheless, the Renyi-2 correlation function for \(\theta^a\) still maintains long-range correlation:
\begin{align}
    \frac{\tr[\rho e^{i\theta^s(y')} e^{-i\theta^s(y)}\rho e^{-i\theta^s(y')} e^{i\theta^s(y)}]}{\tr[\rho^2]}=\frac{1}{|y-y'|^{\eta}}
\end{align}
To provide a physical interpretation of this Rényi-2 correlator, we duplicate our Hilbert space by creating two identical copies of the purified state, denoted \(|\psi^{as}\rangle^{1}\) and \(|\psi^{as,*}\rangle^{2}\). We take the complex conjugate of the second copy of the wave function; its physical interpretation and significance will be explained shortly. These two identical copies, \(|\psi^{as}\rangle^{1} \otimes |\psi^{as,*}\rangle^{2}\), exhibit quasi-long-range order in the four-point correlator, which essentially represents the product of the two-point correlation for each copy of the purified state:
\begin{widetext}
\begin{align}
\langle e^{i(\theta^{a,1}(y)+\theta^{s,1}(y))}e^{-i(\theta^{a,2}(y)+\theta^{s,2}(y))}  e^{-i(\theta^{a,1}(y')+\theta^{s,1}(y'))}e^{i(\theta^{a,2}(y')+\theta^{s,2}(y'))} \rangle= \frac{1}{|y-y'|^{2b}}
\label{stringdual}
\end{align}
\end{widetext}
We now project the i-th ancilla from both the first and second copies, forcing their alignment in the same $\theta$ direction by projecting onto a symmetric EPR pair:

\begin{align}\label{postp}
    \hat{P}_i \sim (\sum_{\theta}|\theta^{a,1} \theta^{a,2} \rangle_i)(\sum_{\theta}
    \langle\theta^{a,1} \theta^{a,2} |_i)
\end{align}

$\hat{P}_i$ is the projection operator for each ancilla pair \(i\). We denote the normalized wavefunction after this projection as \(\Psi_{pp}\). We expect that the post-projection state \(\Psi_{pp}\) exhibits quasi-long-range order in the four-point correlation function:
\begin{widetext}
 \begin{align}
   \langle \Psi_{pp} | e^{\theta^{s,1}(y)} e^{-i\theta^{s,2}(y)}  e^{-i \theta^{s,1}(y')}e^{i\theta^{s,2}(y')} |\Psi_{pp} \rangle=  \frac{\tr[\rho e^{i\theta^s(y)} e^{-i\theta^s(y')}\rho e^{-i\theta^s(y)} e^{i\theta^s(y')}]}{\tr[\rho^2]}=\frac{1}{|y-y'|^{\eta}}
\end{align}
\end{widetext}
Which precisely corresponds to the Renyi-2 correlator. Technically speaking, tracing out the ancilla and obtaining the mixed state density matrix effectively involves projecting the ancilla in both the ket and bra spaces to be identical, essentially aligning them in the same $\theta$ direction.
By considering the bra vector as a duplicate copy, the ancilla tracing procedure is akin to the EPR projection operation in Eq.~\ref{postp}. 
Drawing from this analogy, when calculating the Rényi-\(2\) correlator, the operators acting on the left and right sides of the density matrix \(\rho\) can be interpreted as measuring operators on both copies of purified states after implementing the ancilla EPR projection\cite{sala2024spontaneous}.

To summarize, the intrinsic topological dipole insulator in a mixed-state exhibits a mixed anomaly between strong \(U^e(1)\) and weak \(U^d(1)\) symmetries at the edge. This anomaly is manifested through the edge correlation functions of the mixed-state density matrix. Specifically, the operator charged under \(U^d(1)\) symmetry shows quasi-long-range correlations, while the operator charged under \(U^e(1)\) displays quasi-long-range order in the Renyi-2 correlation but only short-range order in conventional correlation functions.

From a purification perspective, this suggests the presence of an operator carrying \(U^e(1)\) that inherently has long-range order, potentially involving a bound state between the system and ancilla degrees of freedom. When its conventional correlation function is measured in the system's density matrix, the ancilla's contributions are traced out, effectively diminishing the observable long-range order. However, when measuring the Renyi-2 correlator, which is akin to projecting the ancilla into a fixed pattern, the long-range order is still inherited in the post-projected state.

\subsubsection{General discussion on mixed state anomalies}

Finally, we conclude our discussion on edge anomalies by connecting our argument to other studies of mixed-state anomalies in Ref.~\cite{ma2024symmetry,lee2022symmetry,chen2023symmetry,chen2023separability,kawabata2024lieb,hsin2023anomalies,li2023intrinsically,xu2024average}. 
Ref.~\cite{lee2022symmetry,ma2024symmetry} employed Choi-Jamiolkowski formalism to map the density matrix into a bilayer wavefunction, transforming \(\rho = a_{ij} |i\rangle \langle j|\) into \(|\rho\rangle\rangle \sim a_{ij} |i\rangle |j\rangle\). This transformation links the mixed-state anomaly to the quantum anomaly in the Choi-double space and provides a straightforward way to delineate various physical observables and correlation functions, including the Renyi-2 entropy, in the mixed state. 
In our purification argument, we envision a duplicated copy of the purified state. Within this framework, the Choi-doubled state is created by retaining both the original and the duplicated copy of the purified state, and by projecting the ancillae from both copies into an EPR pair. Consequently, the Choi-Jamiolkowski wavefunction corresponds to the post-measurement purified states that are obtained after the projection of the ancillae from both copies.

Additionally, we would like to discuss the generality of our argument concerning other mixed-state anomalies \cite{wang2024anomaly}. Suppose we have a mixed state with weak $G$ symmetry and strong $S$ symmetry, where both $G$ and $S$ are Abelian continuous groups, and the anomaly is perturbative. In this case, the conclusion in Eq.~\ref{edgecor} always holds provided both symmetries are not broken. Specifically, the $G$-charged operators exhibit long-range correlations, while the $S$-charged operators display long-range order in the Renyi-2 correlator. The purification argument in Sec.~\ref{sec:puriano} applies as long as $G$ and $S$ are Abelian continuous groups, ensuring their local density operators are well-defined. However, our argument does not extend to discrete symmetry groups. Ref.~\cite{lessa2024mixed} demonstrates the anomaly in a 1D mixed state with strong $Z_2$ symmetry, whose density matrix is not tripartite-separable. Nonetheless, all operators that are linear in the mixed-state density matrix are short-range correlated. We hope to address these issues more comprehensively in our future work.

\subsection{Hierarchical Structure of Anomalies in Mixed States}\label{sec:dipolestructure}

So far, we have explored the intrinsic topological dipole insulator in open systems, which exhibit a mixed anomaly between weak \(U^d(1)\) and strong \(U^e(1)\) symmetries at the boundary. This raises the question of whether it's possible to achieve an edge pattern that exhibits a self-anomaly regarding the strong \(U^e(1)\) symmetry in mixed-state settings. This scenario is reminiscent of the quantum Hall effect, where a \(U^e(1)\) flux insertion leads to charge pumping between boundaries. A distinctive feature of our scenario, however, is the requirement for the system to maintain weak \(U^d(1)\) symmetry.

We will briefly clarify why such an edge pattern is not feasible. Suppose an anomaly associated with \(U^e(1)\) exists at the edge. In the purification framework, this would manifest as a \(U^e(1)\) current anomaly carried only by the system qubits, with the ancilla remaining neutral under this symmetry. This edge anomaly can be demonstrated by inserting a global flux via a gauge potential \(A_y = \frac{2\pi}{L_y}\), which couples only to the system qubits and induces a charge transfer between the left and right boundaries. Given such patterns at the edge, one can apply a dipole flux to the purified state by introducing a dipole gauge potential \(A^d_y = \frac{2\pi x}{L_y}\). This action results in the left edge (of the system qubits) acquiring a unit of \(U^e(1)\) charge while the right edge loses \(L_x\) units of \(U^e(1)\) charge. Although the ancillae couple to and respond to \(A^d_y = \frac{2\pi x}{L_y}\), they remain neutral under \(U^e(1)\), thus preventing any additional \(U^e(1)\) charge transfer between the edges. Consequently, the total \(U^e(1)\) charge in the purified state changes after the dipole flux insertion, indicating that the purified state has an anomalous bulk that cannot be short-range entangled.

\begin{figure}[h]
    \centering
\includegraphics[width=0.47\textwidth]{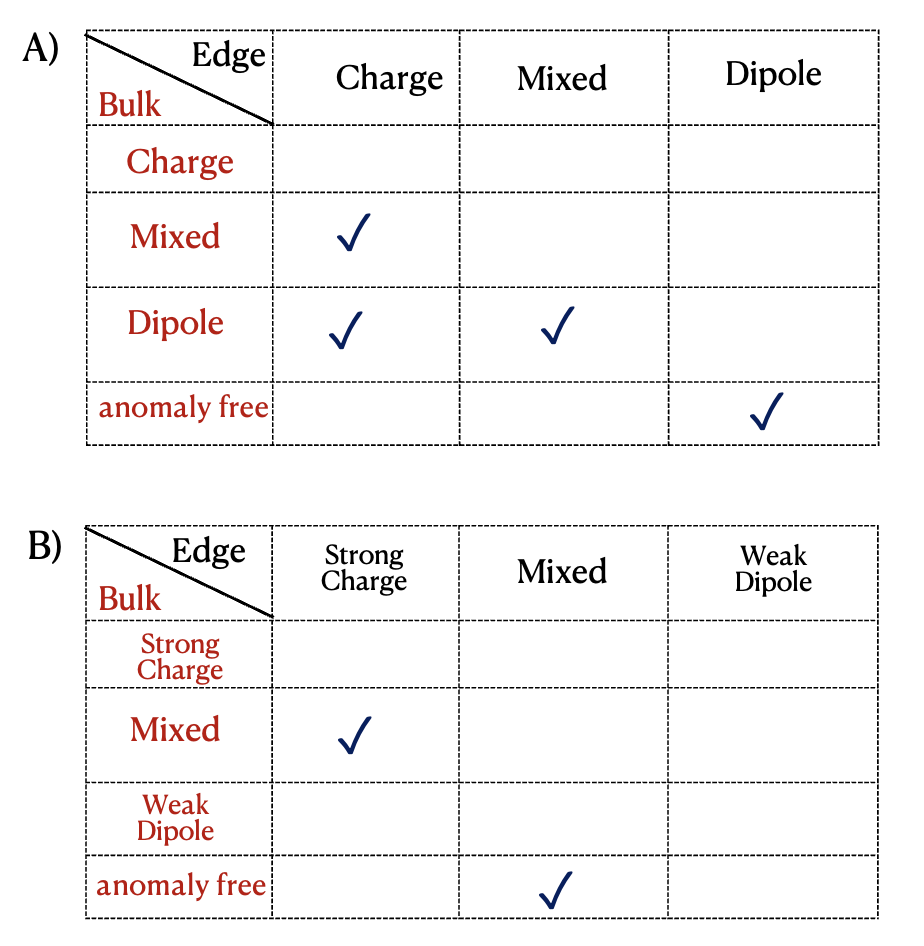}
    \caption{A) Hierarchical Structure of \(U^e(1)\) and \(U^d(1)\) Anomalies in a Closed System.
B) Hierarchical Structure of Strong \(U^e(1)\) and Weak \(U^d(1)\) Anomalies in an Open System.}
    \label{anomalyfig}
\end{figure}

Based on our exploration so far, the 2D SPT phase with charge multipole conservation exhibits a hierarchical structure regarding quantum anomalies\cite{lam2024topological,may2022interaction,zhang2023classification}, as shown in Fig.~\ref{anomalyfig}. 

Let's proceed with systems that conserve both dipole (\(U^d(1)\)) and charge (\(U^e(1)\)) symmetries, initially focusing on a closed system.

a) A mixed anomaly between \(U^d(1)\) and \(U^e(1)\) symmetry at the edge implies that the bulk also has a \(U^d(1)\) anomaly.

b) Similarly, if the boundary has a \(U^e(1)\) anomaly, the insertion of a \(U^e(1)\) flux would alter the dipole moment in the bulk, indicating that the bulk has a mixed anomaly between \(U^d(1)\) and \(U^e(1)\) symmetry. Neither of these can be rendered as the ground state of a local Hamiltonian in 2D.

When considering mixed states in open systems, treating \(U^d(1)\) as a weak symmetry alters the hierarchical structure. In an open quantum system, the system's qubits interact with the ancilla from the environment, permitting fluctuations and exchanges of dipole charges between them. We can arrange the ancilla to carry the opposite `dipole anomaly' in the bulk, rendering the entire purified state anomaly-free. Consequently, a mixed anomaly between weak \(U^d(1)\) and strong \(U^e(1)\) symmetry at the boundary does not induce any anomaly in the bulk.

\section{Intrinsic 3D HOTI in open system}
\label{sec:3dhoti}
Building on our investigations, we have identified an intrinsic 2D topological dipole insulator in a mixed state that lacks an equilibrium analog. The essence of this intrinsic mixed-state symmetry-protected topological (imSPT) phase is that in a pure state—as the ground state of a local Hamiltonian— some edge patterns and boundary anomalies would imply an anomalous bulk. However, in an open quantum system, if certain symmetries are demoted to weak symmetries within a mixed ensemble, the bulk anomaly can be cancelled by the ancilla, rendering the mixed-state density matrix short-range entangled. In the subsequent section, we will expand on this concept and present another example of intrinsic mixed-state SPT with subsystem symmetries.

\subsection{3D HOTI with subsystem symmetries: Absence of chiral hinge state}\label{sec:3dhotinogo}

We turn to another example — an intrinsic 3D higher-order topological insulator (HOTI) in an open system. This example is inspired by the three-dimensional subsystem symmetry-protected higher-order topological insulator (HOTI) previously investigated in a closed system in Ref.~\cite{may2022interaction,zhang2023classification}. Consider a charged insulator defined on a 3D cubic lattice, whose charge is individually conserved on each xz and yz plane. Consequently, the system exhibits a 2-foliated subsystem U(1) symmetry, represented as \(U^{xz}(1)\) and \(U^{yz}(1)\). Ref.~\cite{may2022interaction,zhang2023classification} articulates a no-go theorem to preclude any mixed anomaly between \(U^{xz}(1)\) and \(U^{yz}(1)\) on the 1D hinge (along z-direction), provided the bulk degree of freedom is gapped. It was promptly noted that the presence of such 2-foliated subsystem U(1) symmetry makes it impossible to manifest higher-order topological insulator patterns that feature both gapped bulk and side surfaces, along with chiral modes localized at the hinges in the z-direction.

To illustrate this, we assume that chiral modes exist at four hinges along the z-direction, arranged as follows:
\begin{align}
[\psi^1_{L}(0,0,z), \psi^{2}_{R}(0,L,z), \psi^{3}_{R}(L,0,z), \psi^4_{L}(L,L,z)] \label{hingea}.
\end{align}
\(\psi_{L/R}\) denotes a left/right moving mode along the z-direction.
From now on, we will consider systems placed on a geometry that is periodic along the z-axis and has open boundaries on the x-y plane, with dimensions \(L \times L\) as Fig.~\ref{3D1}.

Given that charge is conserved on each x-z plane, we can apply a subsystem \(U^{xz}(1)\) flux exclusively to the left side surface at \(y=0\). This is achieved by generating a gauge potential \(A_z = \frac{2\pi \delta(y)}{L_z}\), localized at the \(y=0\) side surface. Provided both the bulk and side surface degrees of freedom are gapped, only the two gapless modes \(\psi^1_{L}(0,0,z)\) and \(\psi^3_{R}(L,0,z)\) on the two hinges will react to the subsystem flux insertion, resulting in a charge density shift on the hinges by \(\pm 1\). After the flux insertion, the hinge located at \(x=0, y=0\) transfers charge to the other hinge at \(x=L, y=0\). This mechanism facilitates the creation of an additional charge on the \(x=L\) plane, effectively balancing the charge lost from the \(x=0\) plane.
However, recalling that subsystem \(U^{yz}(1)\) symmetry charge is conserved on all y-z planes, the total charge on each \(x=L\) (or \(x=0\)) plane needs to remain invariant after the flux insertion. The assumption of a chiral hinge state presents an obstruction. It suggests that a large gauge transformation of \(U^{xz}(1)\) symmetry would break the \(U^{yz}(1)\) symmetry for the whole 3D system, leading to an anomalous bulk state. This phenomenon can be described as a \textit{conflict of anomaly cancellation}. For instance, a chiral mode situated at one of the hinges, such as at \(x=0, y=0\), exhibits a mixed hinge anomaly that involves both \(U^{xz}(1)\) and \(U^{yz}(1)\) symmetries. However, given that these are subsystem symmetries with charge conservation on each x-z and y-z plane, there is no self-consistent way to assign anomalous patterns to the remaining hinges to cancel this anomaly. Based on this analysis, we conclude that it is impossible to have a chiral hinge mode for a higher-order topological insulator (HOTI) that adheres to subsystem charge conservations.

 \subsection{Intrinsic 3D HOTI in open systems}

\begin{figure}[h]
    \centering
\includegraphics[width=0.5\textwidth]{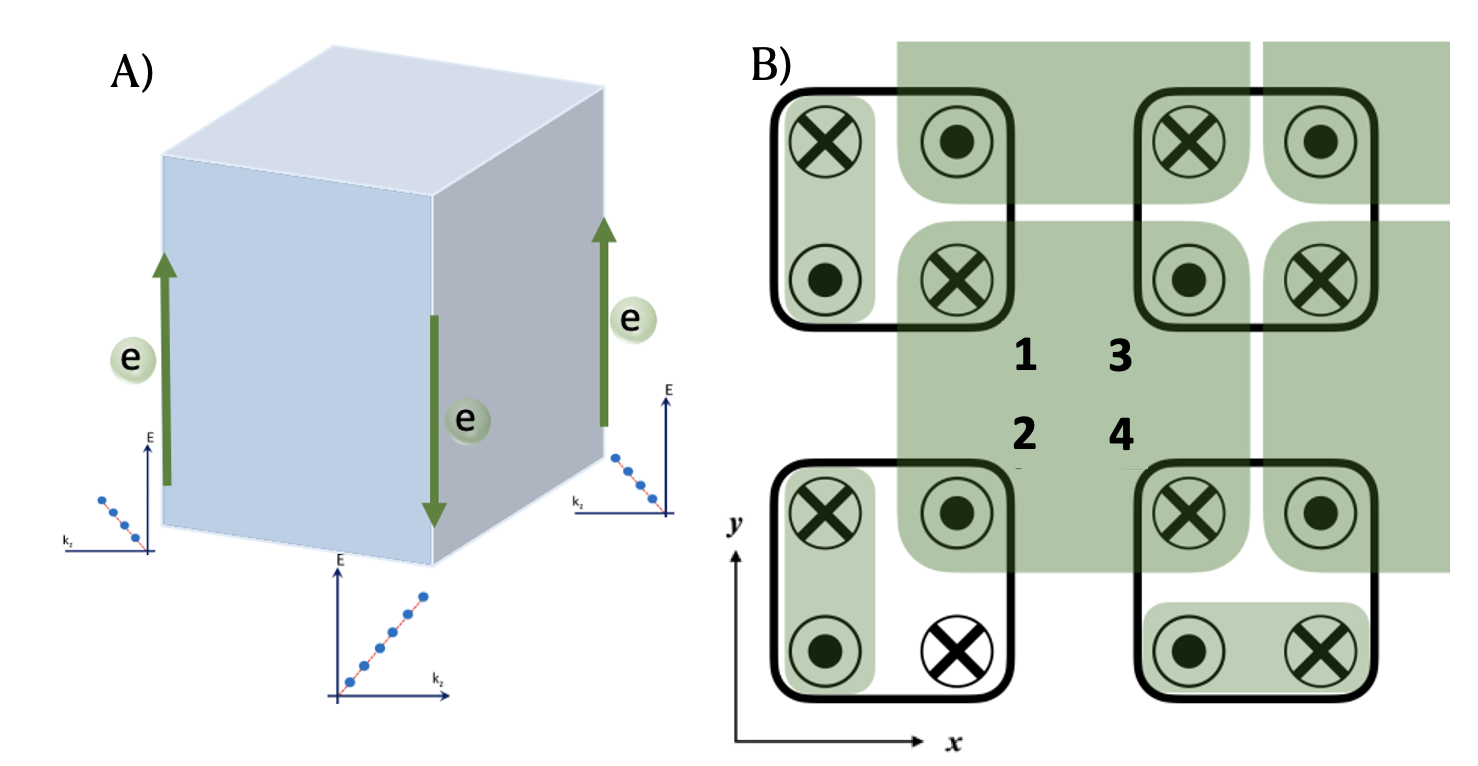}
    \caption{A) Chiral modes localized on the hinge along the z-axis. B) 3D array of 1D wires aligned along the z-direction. Each unit cell (solid-line square) contains two flavors of 1D Luttinger liquid per wire. The elementary building block (green-shaded square) comprises 1D wires from four unit cells at the hinges of the tube.}
    \label{3D1}
\end{figure}

 Now consider HOTI settings in open quantum systems where both \(U^{xz}(1)\) and \(U^{yz}(1)\) symmetries are treated as \textit{weak symmetries}. Meanwhile, global U(1) charge conservation remains a \textit{strong symmetry}, ensuring no charge transfer between the system and the environment.
To streamline the formulation from the coupled wire model introduced in Ref.~\cite{may2022interaction,zhang2023classification,zhang2023fractonic}, we consider a 3D array of 1D wires aligned along the z-direction. Each unit cell contains two flavors of 1D Luttinger liquid per wire. We label these modes as chiral boson fields \(\phi^a_{L/R}(r)\) (where \(a=1,2\)). The elementary building block takes the shape of a thin tube extended along the z-direction, comprising 1D wires from four unit cells at the hinges of the tube, as illustrated in Fig.~\ref{3D1}.
The wires within each building block are:
\begin{align}
\phi^1_{L}(r),~ \phi^2_{R}(r+e_x), ~\phi^1_{R}(r+e_y), ~\phi^2_{L}(r+e_x+e_y)
\end{align}
We first couple these wires within each building block using quartic inter-wire interactions that respect both \(U^{xz}(1)\) and \(U^{yz}(1)\) symmetries:
\begin{align}\label{eq:block}
V_{int} = &v_0 \cos(\phi^{1}_{L}(r)-\phi^{2}_{R}(r+e_x)-\phi^{1}_{R}(r+e_y)\nonumber\\
&+\phi^{2}_{L}(r+e_x+e_y))  
\end{align}
The subsystem symmetries require that the charge within each \(xz\) and \(yz\) plane be independently conserved, thereby excluding any other coupling terms within the building block. As a result, the interaction term \(V_{int}\) only gaps out one of the gapless branches in the building block, while the other remains gapless.

To simplify the notation, we express the chiral boson fields in each building block in terms of chiral fermion operators as \(\psi_{R/L} \sim e^{i\phi_{R/L}}\). In this representation, each building block is equivalent to a four-component Dirac spinor:
\begin{align}
&\mathcal{H}_{\text{wires}}=\sum_{\bm{r}} \bm{\psi}^\dagger_{\bm{r}} i\partial_z \tau^{zz}\bm{\psi}_{\bm{r}},\nonumber\\
    &\psi^{\dagger}=(\psi^{\dagger}_{L,1},\psi^{\dagger}_{R,2},\psi^{\dagger}_{R,3},\psi^{\dagger}_{L,4}),\nonumber\\
    &\psi_{L,1}=e^{ i \phi^{1}_{L}(r)},
       \psi_{R,2}=e^{ i \phi^{2}_{R}(r+e_x)},\nonumber\\
          &\psi_{R,3}=e^{ i \phi^{1}_{R}(r+e_y)},
             \psi_{L,4}=e^{ i \phi^{2}_{L}(r+e_x+e_y)}
    \label{hinge}
    \end{align} 
Here, and throughout, we use the shorthand \(\tau^{ij...k} \equiv \tau^i \otimes \tau^j \otimes...\otimes \tau^k\), where \(\tau^i\) are the Pauli matrices.
The subsystem symmetry acts on the four-component Dirac spinor as follows:
\begin{align}
     U^{xz}(1): &   \psi^{\dagger} \rightarrow  e^{i \alpha \tau^{z0}}\psi^{\dagger}\nonumber\\
       U^{yz}(1): &   \psi^{\dagger} \rightarrow  e^{i \beta \tau^{0z}}\psi^{\dagger}
\end{align}
Based on this symmetry assignment, no fermion bilinear mass can be added to the building block.

We now couple the Dirac spinor that represents the four wires in each building block to a fluctuating O(4) mass vector $\vec{m}(r) = (m_{1},m_{2},m_{3},m_4)$:
\begin{equation}\begin{split}
H=\psi^{\dagger}\Big[&i\partial_z \tau^{zz}+m_{1}\tau^{xz}+m_{2}\tau^{yz}\\
&+m_{3}\tau^{0x}+m_{4} \tau^{0y}\Big]\psi.
\label{eq:hinge}\end{split}\end{equation}
We can interpret these mass vectors as dynamical masses fluctuating in space-time, generated through the interaction term \(V_{int}\).
Triggered by strong O(4) mass fluctuations, the fermion excitations in each building block, as described in Eq.~\ref{eq:hinge}, become massive due to dynamical mass generation. Consequently, we can integrate out the fermions and obtain an effective theory for the O(4) mass vector. This effective theory is akin to the non-linear sigma model with a Wess-Zumino-Witten (WZW) term:
\begin{equation}\begin{split}
&\mathcal{L}=\frac{1}{g}(\partial_{\mu} \vec{m})^2+\frac{2\pi}{\Omega^3} \int_0^1 du\epsilon^{ijkl}  m_{i}\partial_z m_{j} \partial_t m_{k}\partial_u m_{l},
\end{split}\label{wzw}
\end{equation}
As a result, the fermion excitations are gapped in each building block, while a collective gapless mode with bosonic excitations remains.
The subsystem symmetry \(\text{U}(1)^{xz}\) rotates between \(m_{3}\) and \(m_{4}\), while \(\text{U}(1)^{yz}\) rotates between \(m_{1}\) and \(m_{2}\). Thus, the WZW term in Eq.~\ref{wzw} implies a mixed anomaly between the \(\text{U}(1)^{xz} \times \text{U}(1)^{yz}\) symmetry\cite{xu2013wave, bi2015classification,you2016stripe}. The physical effect of this WZW is that a \(2\pi\) flux insertion for \(\text{U}(1)^{xz}\) would trigger a charge shift for \(\text{U}(1)^{yz}\).

Now consider adding quenched disorder to each building block by introducing a disordered mass vector \(\vec{m}(r,t)\). Although each specific disorder mass pattern explicitly breaks subsystem symmetry, when we consider mixed ensembles of all possible disorder mass configurations, the resulting mixed-state density matrix \(\rho\) still exhibits a weak \(\text{U}(1)^{xz} \times \text{U}(1)^{yz}\) symmetry. Likewise, since the vector \(\vec{m}(r,t)\) is neutral under global U(1) symmetry, the density matrix also respects strong U(1) symmetry. The mixed state density matrix has the following form:
\begin{align}\label{mixed0}
    \rho=\sum_{\{\vec{m} \}} |\vec{m}\rangle \langle \vec{m} |
\end{align}
For each specific vector mass pattern \(\vec{m}\), the ket vector \(|\vec{m}\rangle\) denotes the ground state resulting from the static vector mass \(\vec{m}\) coupling to the fermions, as described in Eq.~\ref{eq:hinge}, within each building block. In the mixed ensemble, the Berry phase effect triggered by the Wess-Zumino-Witten term is canceled by the quenched disorder, leading to a short-range entangled density matrix for each building block. Thus, the quenched disorder induces a short-range correlated mixed state throughout the bulk.

What happens to the boundaries? For smooth boundaries on the side surfaces, each unit cell features a pair of up/down moving modes that are dangling and decoupled from any other building block in the bulk, illustrated as Fig.~\ref{3D1}. Since they are located in the same unit cell and have the same position index, we can gap them out via intra-wire coupling.
For rough boundaries, such as the hinge along the z-direction, there is an additional chiral mode that remains decoupled from any settings. This chiral mode cannot be trivialized into a short-range entangled mixed state by quenched disorder or infinite-depth quantum channels in an open system setting. 
From the anomaly aspect, if we insert a subsystem \( U^{xz}(1) \) flux at the left side surface, it will trigger a charge pumping between the top left and bottom left hinges, indicating that each hinge manifests a mixing anomaly between the weak \( U^{xz}(1) \) and strong \( U(1) \) symmetries. Notably, such a `chiral current' on the hinge cannot manifest in a closed system, as it infers a mixed anomaly between the \( U^{xz}(1) \) and \( U^{yz}(1) \) symmetries in the bulk. In an open system, if we treat `subsystem symmetries' as weak symmetries, the subsystem charge can exchange and fluctuate with the ancilla from the environment. As a result, the obstruction of the subsystem symmetry anomaly from the bulk can potentially be canceled by the ancilla.

\section{Outlook}
In this work, we display a path to search for intrinsic mixed-state SPT by considering open systems with modulated charge conservation, such as dipole moment or subsystem charge. The key idea is that a system with modulated symmetry can display a hierarchical structure for the quantum anomaly, provided that the global symmetry and modulated symmetry groups are intertwined rather than forming a direct product. When contemplating mixed states in open systems with \textit{weak modulated symmetry}, it is equivalent to considering a purified state with additional ancillae from the environment that also carries modulated symmetry. Thus, some obstructions can be eliminated, and the hierarchical anomaly structure is modified. We conclude our discussion by outlining some future directions:

1) In this study, we explore intrinsic mixed-state symmetry-protected topological (imSPT) phases featuring subsystem or dipole symmetries. There is a pressing need for extensive exploration of a broader class of modulated symmetry SPT in open systems, such as those exhibiting incommensurate\cite{sala2022dynamics,sala2024exotic}, fractal\cite{zhou2021fractal}, or exponential symmetry\cite{delfino20232d,watanabe}.

2) We identify edge anomalies in mixed-state SPTs by analyzing edge correlation functions and Renyi-2 correlations. However, our analysis is limited to perturbative anomalies within continuous symmetry groups. In the context of discrete symmetries, it is observed that the correlation function\cite{lessa2024mixed,xu2024average} in mixed states can exhibit short-range characteristics. The nature of mixed-state anomalies still demands further investigation. Additionally, Ref.~\cite{lessa2024mixed,chen2023separability} explores the characteristics of the separability condition of mixed-state density matrices in open-system SPTs. It would be valuable to examine how these conditions apply to our intrinsic mixed-state SPTs.

3) Ref.~\cite{lam2024topological,huang2023chern} introduces a dipolar Chern-Simons response theory to characterize the topological linear response of TDI in closed systems. The potential for a field theory\cite{ma2024symmetry,ellison2024towards,sohal2024noisy,bao2023mixed} description of open SPTs under the Keldysh formalism is deferred for future exploration.

\acknowledgements 
We thank Jong-Yeon Lee, Pablo Sala, Zhen Bi, Ho Tat Lam, and Ruochen Ma for their helpful discussions and comments. In particular, we acknowledge Chong Wang for pointing out the use of the algebra of current operators to demonstrate long-range correlation functions in mixed states.
This work was performed in part at the Aspen Center for Physics (MO, YY), which is supported by the National Science Foundation grant PHY-2210452 and Durand Fund(YY). This research was supported in part by grant NSF PHY-2309135 to the Kavli Institute for Theoretical Physics (KITP).
M. O. was partially supported by JSPS KAKENHI Grant No. JP24H00946.

\end{document}